\documentclass[english,pre, twocolumn]{revtex4-1}
\usepackage[T1]{fontenc}
\usepackage[latin9]{inputenc}
\usepackage{amstext}
\usepackage{amssymb}
\usepackage{graphicx}
\usepackage{esint}
\usepackage{babel}
\begin{document}

\title{Fisher Waves: an individual based stochastic model.}

\author{B. Houchmandzadeh and M. Vallade}

\affiliation{CNRS, LIPHY, F-38000 Grenoble, France~\\
Univ. Grenoble Alpes, LIPHY, F-38000 Grenoble, France}
\begin{abstract}
The propagation of a beneficial mutation in a spatially extended population
is usually studied using the phenomenological stochastic Fisher-Kolmogorov
(SFKPP) equation. We derive here an individual based, stochastic model
founded on the spatial Moran process where fluctuations are treated
exactly. At high selection pressure, the results of this model are
different from the classical FKPP. At small selection pressure, the
front behavior can be mapped into a Brownian motion with drift, the
properties of which can be derived from microscopic parameters of
the Moran model. Finally, we show that the diffusion coefficient and
the noise amplitude of SFKPP are not independent parameters but are
both determined by the dispersal kernel of individuals. 
\end{abstract}
\maketitle

\section{Introduction.}

One of the most fundamental questions in evolutionary biology is the
spread of a mutant with fitness $1+s$ into a wild type population
with fitness $1$. In a non-structured population (\emph{i.e.,} for
a population at dimension $d=0$), the answer to this question was
found by Kimura\cite{Kimura1962} nearly 50 years ago as a good approximate
solution of the Fisher-Wright or the Moran model of population genetics,
and better solutions of the Moran model have been proposed recently\cite{Houchmandzadeh2010}.
For geographically structured populations however, the question is
far from settled and only some specific information, such as the fixation
probability, has received partial answers in a field that is now called
evolutionary graph dynamics\cite{Lieberman2005,Houchmandzadeh2011b}.
For geographically structured populations where the main ingredients
of the competing populations, \emph{i.e., }the fitness, the carrying
capacity and the diffusion of individuals, are independent of the
space, the evolutionary dynamics has been mostly investigated through
the stochastic Fisher Kolmogorov Petrovsky, Piscounov (SFKPP) equation
\begin{equation}
\frac{\partial u}{\partial t}=D\nabla^{2}u+au(1-u)+\sqrt{bu(1-u)}\eta(x,t)\label{eq:FKPP}
\end{equation}
where $u(x,t)$ is the local relative density of the mutant with respect
to the local carrying capacity, $D$ is the diffusion coefficient
of individuals, $a$ is proportional to the relative excess fitness
of mutants; the last term is a noise term that captures the local
genetic drift, where $b$ is related to the local carrying capacity
and $\eta$ is a white noise. The problem that has attracted most
attention is that of the front propagation : if at the initial time,
one half of space is filled only with the mutant type and the other
half only with the wild type, then the dynamics of the problem can
be reduced to the dynamics of the front separating the two types. 

The deterministic part of the equation (FKPP) was proposed by Fisher\cite{Fisher1937}
and Kolmogorov, Petrovsky, Piscounov\cite{Kolmogorov}; it has found
applications in many areas of science ranging from ecology and epidemiology\cite{Murray2007}
to chemical kinetics\cite{Beuer1995} and particle physics\cite{Munier2003}.
The properties of the FKPP equation have been widely investigated\cite{Vansaarloos2003}.
Specifically, this equation allows for traveling wave solutions and
it is known that a stable solution of the FKPP is a wave front connecting
the two regions $u=1$ and $u=0$ with velocity $c=(d/dt)\int_{\mathbb{R}}udx=2\sqrt{aD}$
and width $B=\int_{\mathbb{R}}u(1-u)dx=2\sqrt{D/a}$. 

The FKPP equation however is not well adapted to evolutionary population
dynamics at small selection pressure, which is one of the relevant
limits of population genetics\cite{Kingsolver2001,Nei2005}. The FKPP
equation describes quantities (individuals, molecules,...) that at
the fundamental level are discrete; the noise associated with this
discreteness can play an important role in the dynamics of the front,
specifically at small selection pressures. This problem was tackled
phenomenologically by adding either a cutoff \cite{Brunet1997} or
alternatively a noise term to the equation. The form of the noise
in the SFKPP was proposed by Doering et al.\cite{Doering2003}. The
SFKPP proposed by Doering et al. has now become a major mathematical
tool for the investigation of Fisher Waves. It has specifically been
used by Hallatschek and Korolev\cite{Hallatschek2009} to investigate
the properties of the front at small selection pressure, where they
revealed the marked difference of the solutions with respect to the
deterministic equation.

The SFKPP equation however is phenomenological and cannot be derived
rigorously from a microscopic, individual based model of population
genetics. Firstly, individual based models such as the Moran model
are governed by discrete master equations and can be approximated
by a Fokker-Planck equation, or their equivalent stochastic differential
equation, only in the limit of large system size, \emph{i.e.,} large
local carrying capacity\cite{Ethier1977,Houchmandzadeh2010}. The
local carrying capacity however does not appear explicitly in the
SFKPP equation and it is difficult to assess the precision of the
Focker-Planck approximation solely from this equation. Secondly, and
more importantly, the noise term $\sqrt{bu(1-u)}\eta(x,t)$ in SFKPP
is purely local. This noise term is rigorous only for a 0 dimensional
system, where the equivalence between the Fokker-Planck approximation
and the stochastic differential equation can be shown. For a spatially
extended system, the noise term should also include fluctuations arising
from adjacent lattice cells. To our knowledge, however, a rigorous
derivation has not yet been achieved (see Mathematical Details \ref{subsec:NoiseSFKPP}).
The problem of noise arising from adjacent cells was also noted by
Korolev et al.\cite{Korolev2010}. Finally, in an evolutionary model,
both the diffusion coefficient and the noise amplitude are the result
of the same phenomenon of individuals replacing each other randomly
and they should be linked through an Einstein like relation. 

The aim of the present article is to study the dynamics of the front
between mutant and wild type individuals directly from the individual
based, stochastic Moran model of population genetics. For this model,
the Master equation can be stated without ambiguity or approximation.
We show that the mean field approximation of the Master equation gives
rise to a partial differential equation that differs from the FKPP
equation and its predictions at high selection pressure. Going beyond
mean field, we then derive the exact equations for the evolution of
the various moments of the front for a one dimensional system and
solve it at small selection pressure. In this approach, the noise
term is not restricted to be only local. We show, in agreement with
\cite{Hallatschek2009} that even for a neutral model (\emph{i.e.,
$s=0$}), the front is well defined and the displacement of the front
can be mapped into a Brownian motion at large times, the convergence
time to this state is shown to be in $1/\sqrt{t}$. The front drift
and its velocity can then be derived at small selection pressure by
a perturbatiion approach where we can show, in contrast to the FKPP
predictions, that the speed of the front is linear in the excess relative
fitness $s$. Finally, we show that the \emph{effective }local population
size which controls the noise amplitude, and the diffusion coefficient
are both determined by the dispersal kernel of individuals and cannot
be chosen as independent parameters. 

This article is organized as follow. Section \ref{sec:The-1d-system}
is devoted to the generalization of the Moran model to population
geographically structured into demes/islands, where the dynamics of
the front can be deduced from the internal population dynamics of
the islands and their exchanges. We demonstrate in subsection \ref{subsec:Mean-field-approximation.}
how an FKPP-like equation emerges from the mean field approximation
of the Master equation and show how it differs from the classical
FKPP equation. The following subsections of section \ref{sec:The-1d-system}
are devoted to full stochastic treatments of the dynamics of the front.
Section \ref{sec:Microscopic model.} goes beyond the island model
and considers general migration kernels between individuals that are
no longer grouped into demes. Solving the Master equation of the model
shows how the island size and migration number between neighboring
islands of section \ref{sec:The-1d-system} are related through the
dispersal kernel. The approach allows for the determination of the
effective population size and therefore the noise amplitude. The final
section is devoted to discussion and conclusions. 

\section{The island model and mutants propagation in 1 dimension.\label{sec:The-1d-system}}

The fundamental model of population genetics for non structured populations\cite{Ewens2004}
was formulated by Fisher and Wright\cite{Fisher1999}. A continuous
time version, which is also more mathematically tractable, was proposed
by Moran\cite{Moran1962}. The extension of the Moran model to geographically
extended populations was formulated by Kimura\cite{Kimura1964a} and
Maruyama\cite{Maruyama1974} and in more recent terminology is referred
to as evolutionary dynamics on graphs \cite{Lieberman2005}. The model
is also widely used in ecology, specifically in the framework of the
neutral theory of biodiversity\cite{Hubbel2001,Houchmandzadeh2003,Vallade2003}. 

\begin{figure}
\begin{centering}
\includegraphics[width=0.8\columnwidth]{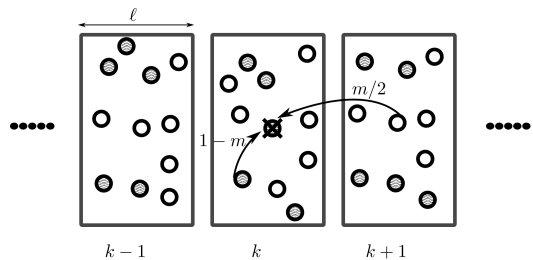}
\par\end{centering}
\caption{The spatial Moran model for geographically extended populations.\label{fig:moranspace}}
\end{figure}

In this model, populations, formed of wild type individuals with fitness
1 and mutants with fitness $1+s$ are structured into cells (or demes
or islands) each containing $N$ individuals. When an individual dies
in one island, it is immediately replaced by the progeny of another,
therefore keeping the number of individuals in each island always
equal to $N$. The replacement probability is weighted by the fitness
of the individuals; moreover, the progeny stems from a local parent
with probability $(1-m)$ or a parent from a neighboring island with
probability $m$ (figure \ref{fig:moranspace}). These three parameters,
$N$, $s$ and $m$ are the only ingredients of this generic model.

Let us consider an infinitely extended one-dimensional collection
of islands and call $n_{i}$ the number of mutant individuals on island
$i$. We use the vector $\mathbf{n}$ as a shorthand notation for
the collection of these numbers $\mathbf{n}=\{...,n_{i},...\}$. The
transition probability densities for the number of mutants on island
$i$ to increase/decrease by one individual is\cite{Houchmandzadeh2011b}:
\begin{eqnarray}
W_{i}^{+}(\mathbf{n}) & = & \frac{\mu(1+s)}{N}(N-n_{i})\left[n_{i}+\frac{m}{2}n''_{i}\right]\label{eq:rateup}\\
W_{i}^{-}(\mathbf{n}) & = & \frac{\mu}{N}n_{i}\left[(N-n_{i})-\frac{m}{2}n''_{i}\right]\label{eq:ratedown}
\end{eqnarray}
where $\mu$ is the death rate of individuals and 
\begin{equation}
n''_{i}=(n_{i-1}+n_{i+1}-2n_{i}).\label{eq:n''def}
\end{equation}
The rate of increase (\ref{eq:rateup}) for example is the probability
density per unit of time that one wild type individual dies ($\mu(N-n_{i})$
) multiplied by the probability that it is replaced by a mutant, either
from a local parent ($(1-m)n_{i}/N$) or a neighboring parent ($(m/2N)(n_{i-1}+n_{i+1})$
), and multiplied by the fitness of the mutant $(1+s)$. The fitness
can be seen as an increase in the death rate of the wild type individuals,
or a higher replacement probability/decreased death rate for the mutants.
The probability $P(\mathbf{n},t)$ of observing the state $\mathbf{n}$
at time $t$ obeys the Master equation 
\begin{eqnarray}
\frac{dP(\mathbf{n})}{dt} & = & \sum_{i}\left(W_{i}^{+}(a_{i}\mathbf{n})P(a_{i}\mathbf{n})-W_{i}^{+}(\mathbf{n})P(\mathbf{n})\right)\nonumber \\
 & + & \left(W_{i}^{-}(a_{i}^{\dagger}\mathbf{n})P(a_{i}^{\dagger}\mathbf{n})-W_{i}^{-}(\mathbf{n})P(\mathbf{n})\right)\label{eq:MasterEquation}
\end{eqnarray}
where $a_{i}^{\dagger}\mathbf{n}$ and $a_{i}\mathbf{n}$ are shorthand
notations for states $\{...,n_{i}\pm1,...\}$. Without loss of generality
(see Mathematical Details \ref{subsec:Choice-of-initial}), the initial
condition we use throughout this article is that of an initial sharp
front
\begin{equation}
n_{i}=N\,\,\,\mbox{if}\,\,i\le0;\,\,=0\,\,\mbox{otherwise}\label{eq:sharpfront}
\end{equation}

\subsection{Mean field approximation.\label{subsec:Mean-field-approximation.}}

\begin{figure}
\begin{centering}
\includegraphics[width=0.9\columnwidth]{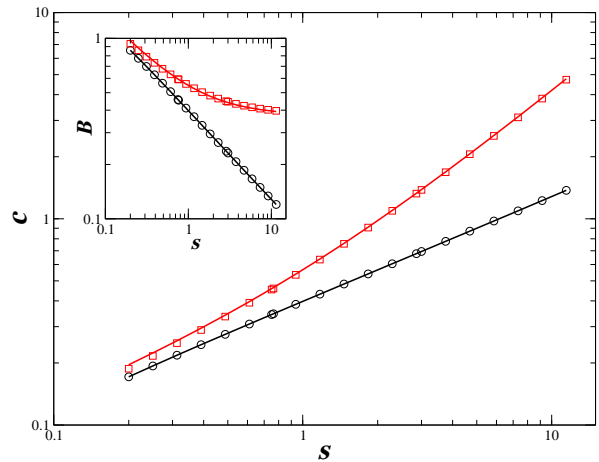}
\par\end{centering}
\caption{The FKPP (black circle) and the spatial Moran mean field (equation
\ref{eq:meanfield}) (red squares) are solved numerically for $\mu=1,$
$D=0.04$ and the front speed is extracted for various values of the
excess relative fitness $s$. Solid curves represent the theoretical
values: FKPP $c=2\sqrt{\mu Ds}$; Moran $c=2\sqrt{\mu Ds(1+s)}$.
Inset : Width of the front as a function of $s$. Black circle: FKPP;
red square: Moran. Solid curves represent, for FKPP (black) $B=2\sqrt{D/\mu s}$;
for Moran (red) $B=2\sqrt{D(1+s)/\mu s}\,\kappa(s)$ (eq. \ref{eq:meanwidth}).
\label{fig:Frontspeed} }
\end{figure}

The mean field approximation for $\left\langle n_{i}\right\rangle $,
the average number of individuals on island $i$, is obtained by neglecting
fluctuations (\emph{i.e., }by setting $\left\langle n_{i}n_{k}\right\rangle =\left\langle n_{i}\right\rangle \left\langle n_{k}\right\rangle $
)
\begin{eqnarray*}
\frac{d\left\langle n_{i}\right\rangle }{dt} & = & \left\langle W_{i}^{+}(\mathbf{n})-W_{i}^{-}(\mathbf{n})\right\rangle \\
 & = & \frac{\mu m}{2}\left\langle n''_{i}\right\rangle +\frac{s\mu}{N}\left(N-\left\langle n_{i}\right\rangle \right)\left(\left\langle n_{i}\right\rangle +\frac{m}{2}\left\langle n''_{i}\right\rangle \right)
\end{eqnarray*}
Taking the space continuum limit by setting $x=\ell i$, $u(x)=n_{i}/N$,
we obtain the partial differential equation 
\begin{equation}
\frac{\partial u}{\partial t}=D\left[1+s(1-u)\right]\frac{\partial^{2}u}{\partial x^{2}}+\mu su(1-u)\label{eq:meanfield}
\end{equation}
where the length $\ell$ is the spatial extension of an island and
$D=\mu m\ell^{2}/2$ is the diffusion coefficient. We observe that
the mean field equation of the spatial Moran model is different from
the FKPP equation in the diffusion term. Fisher himself, in his original
article\cite{Fisher1937}, had stressed that using a simple diffusion
term is an oversimplification of basic population genetics processes.
The modification of the diffusion term has important consequences
both on the speed of the propagation front and on its width. The minimum
speed of the propagating wave in the equation (\ref{eq:meanfield})
is now (see Mathematical details \ref{subsec:meanfieldspeed})
\begin{equation}
c_{min}=2\sqrt{\mu Ds(1+s)}\label{eq:speed_meanfield}
\end{equation}
which scales as $s$ for high value of the excess fitness, in contrast
to the scaling in $\sqrt{s}$ in the FKPP equation. Numerical resolutions
of eq.(\ref{eq:meanfield}) (Figure \ref{fig:Frontspeed}) show that
$v_{min}$ computed above is an excellent estimator of the speed of
the front.
Furthermore, the width of the front does not scale as $\sqrt{D/\mu s}$
as in the case of the FKPP equation, but is well approximated by 
\begin{equation}
2\sqrt{D(1+s)/\mu s}\,\kappa(s)\label{eq:meanfield-width}
\end{equation}
where $\kappa(s)$ is a small correction (see Mathematical details
\ref{subsec:meanfieldspeed}). Specifically, for large $s$, the width
converges to a constant $(15/8)\sqrt{D/\mu}$. 

A phenomenological argument can be used to understand these modifications
to the FKPP equation. It is well known\cite{Panja2003} that the dynamics
of a pulled front is governed by the behavior at small $u$. In these
regions, the mean field equation (\ref{eq:meanfield}) can indeed
be approximated by an FKPP equation, with the effective diffusion
coefficient $D_{eff}=D(1+s)$. 

We observe that the spatial Moran model differs significantly from
the prediction of FKPP equation at high fitness $s$. We will show
below that the same is true at low fitness. This difference was first
noted by Hallatschek and Korolev\cite{Hallatschek2009} in their study
of the SFKPP equation. 

\subsection{Stochastic characterization of the front.\label{subsec:Stochastic-characterization}}

Let us now come back to the full stochastic treatment of the propagating
front. The temporal evolution of local population moments can be extracted
from the Master equation (\ref{eq:MasterEquation}) (see Appendix
\ref{subsec:Moment-computation-algebra.}): 
\begin{eqnarray*}
\frac{d\left\langle n_{i}\right\rangle }{dt} & = & \left\langle W_{i}^{+}-W_{i}^{-}\right\rangle \\
\frac{d\left\langle n_{i}n_{j}\right\rangle }{dt} & = & \left\langle n_{i}\left(W_{j}^{+}-W_{j}^{-}\right)+n_{j}\left(W_{i}^{+}-W_{i}^{-}\right)\right\rangle \\
 & + & \delta_{i,j}\left\langle W_{i}^{+}+W_{i}^{-}\right\rangle .
\end{eqnarray*}
The most important \emph{global} quantities are the front displacement
$U(t)$ and its width $B(t)$. These global quantities can be measured
in terms of local populations by 
\begin{eqnarray}
U(t) & = & \frac{1}{N}\sum_{i=-\infty}^{+\infty}\left[n_{i}(t)-n_{i}(0)\right]\label{eq:displacement}\\
B(t) & = & \frac{1}{N^{2}}\sum_{i=-\infty}^{+\infty}n_{i}(t)\left[N-n_{i}(t)\right]\label{eq:width}
\end{eqnarray}
The width $B(t)$ weights the region where the mutant population
is different from either $0$ or $N$, and in the continuous limit,
can be expressed as $B=\int_{I}u(1-u)dx$ where $u(x)=n_{i}/N$. Note
that these quantities are always finite, as the sums involve only
a finite number of non-zero terms. We restrict this paper to the computation
of the first moments of these quantities, namely $\left\langle U(t)\right\rangle $,
$\mbox{Var}(U(t))$ and $\left\langle B(t)\right\rangle $, where
$\left\langle \right\rangle $ stands for ensemble average. The computation
of these quantities implies the computation of the second moments
\begin{equation}
Z_{p}(t)=\frac{1}{N^{2}}\sum_{i}\left\{ \left\langle n_{i}(t)n_{i+p}(t)\right\rangle -n_{i}(0)n_{i+p}(0)\right\} \label{eq:Zp}
\end{equation}
The width of the front is then 
\begin{equation}
\left\langle B(t)\right\rangle =-Z_{0}(t)+\left\langle U(t)\right\rangle +B(0)\label{eq:Bt}
\end{equation}
and the variance of its displacement is 
\begin{eqnarray}
V(t) & = & \left\langle U^{2}(t)\right\rangle -\left\langle U(t)\right\rangle ^{2}\nonumber \\
 & = & \frac{1}{N^{2}}\sum_{i,j}\left\langle n_{i}n_{j}\right\rangle -\left\langle n_{i}\right\rangle \left\langle n_{j}\right\rangle \label{eq:vartiancedef}
\end{eqnarray}
For the neutral front ($s=0$), self consistent, exact equations without
any moment closure approximation can be derived directly for the global
quantities. At small selection pressures $Ns\ll1$, they can be recovered
through a first order perturbation analysis. 

\subsection{Behavior of the neutral front $s=0$.}

For a one dimensional system where mutants and wild type have the
same fitness ($s=0$), we will show that the front separating these
two populations can be envisioned as a well defined object that performs
a Brownian motion and whose width fluctuates around an equilibrium
value: $\left\langle U(t)\right\rangle =0$ and for large times, $V(t)=mt$
and $\left\langle B(t)\right\rangle =m(N-1)/2$. 

To obtain the above quantities, we sum over local fluctuations
\begin{eqnarray*}
\frac{1}{\mu}\frac{d\left\langle n_{i}\right\rangle }{dt} & = & \frac{m}{2}\left\langle n''_{i}\right\rangle \\
\frac{1}{\mu}\frac{d\left\langle n_{i}n_{j}\right\rangle }{dt} & = & \frac{m}{2}\left(\left\langle n''_{i}n_{j}\right\rangle +\left\langle n_{i}n''_{j}\right\rangle \right)\\
 & + & \delta_{i,j}\left\{ \frac{2}{N}\left\langle n_{i}(N-n_{i})\right\rangle +\frac{m}{2}\left\langle n''_{i}\right\rangle -\frac{m}{N}\left\langle n_{i}n''_{i}\right\rangle \right\} 
\end{eqnarray*}
There are different contributions to $d\left\langle n_{i}n_{i}\right\rangle /dt$
: one is the \emph{local} demographic noise $2\left\langle n_{i}(N-n_{i})\right\rangle /N$,
which appears in the SFKPP equation ; the other term, $m\left\langle n_{i}n''_{i}\right\rangle /N$,
is the demographic noise due to adjacent cells and cannot \emph{a
priori} be neglected\emph{.} In the extreme case where $N=1$ and
therefore $m=1$, the \emph{local} demographic noise is exactly zero,
but the stochasticity of the system remains the same, as we will see
below.

From now on, we will measure time in generation time units, \emph{i.e.,}
set $t\leftarrow\mu t$. By summing over the first moments, we find
trivially that the mean front position stays at its initial value
\[
\frac{d}{dt}\left\langle U(t)\right\rangle =\frac{m}{2}\sum_{i}\left\langle n''_{i}\right\rangle =0
\]
The second moments on the other hand obey a set of linear differential
equations
\begin{eqnarray}
\frac{1}{m}\frac{dZ_{0}}{dt} & = & -2(1+\alpha)Z_{0}+2(1-\beta)Z_{1}+(1-\beta)C_{0}\label{eq:Z0}\\
\frac{1}{m}\frac{dZ_{p}}{dt} & = & -2Z_{p}+Z_{p+1}+Z_{p-1}+C_{p}\,\,\,\,\,\,p>0\label{eq:Zk}
\end{eqnarray}
where $\alpha=(1-m)/(Nm)$, $\beta=1/N$ and the coefficients $C_{p}$
depend on the initial conditions :
\[
C_{p}=\frac{1}{N^{2}}\sum_{i}n_{i}(0)n''_{p+i}(0)
\]
The parameters $\alpha$ and $\beta$ measure the relative contribution
to the demographic noise of local versus adjacent cells . The parameter
$\beta$ can be neglected with respect to $\alpha$ only in the limit
of small migration probability $m\ll1$. 

For an initially sharp front (eq. \ref{eq:sharpfront}), $C_{p}=-\delta_{p,0}$.
We stress that we can assume this condition without loss of generality
(see Appendix \ref{subsec:Choice-of-initial})

The above system (\ref{eq:Z0},\ref{eq:Zk}) can be solved\cite{Houchmandzadeh2003}
exactly. In the Laplace space where $\hat{Z}_{p}(\omega)=\int_{\mathbb{R}^{+}}\exp(-\omega t)Z_{p}(t)dt$,
the solution is particularly simple, 
\[
\hat{Z}_{p}(\omega)=\frac{(1-\beta)C_{0}}{\omega}\frac{z}{z^{2}+2\alpha z+2\beta-1}\frac{1}{z^{p}}
\]
where $\omega=(z+1/z)-2$. By taking the inverse Laplace transform
the exact solution of $Z_{p}(t)$ can be found as a combination of
modified Bessel functions\cite{Houchmandzadeh2003}. In this article,
we are mostly concerned with the large time limit, which can be deduced
from the expansion of $\hat{Z}_{p}(\omega)$ around $\omega=0$ :
\begin{eqnarray}
\hat{Z}_{p}(\omega) & = & -\frac{m(N-1)}{2}\left(\frac{1}{\omega}+\frac{p-K}{\sqrt{\omega}}\right)+\mathcal{O}(1)\nonumber \\
Z_{p}(t) & = & -\frac{m(N-1)}{2}\left(1+\frac{p-K}{\sqrt{\pi mt}}\right)+o(t^{-1/2})\label{eq:Znt}
\end{eqnarray}
where $K=m(N-1)+2$. The above approximation is valid for $t\gg p^{2}$;
a uniform large time approximation for all $p$ can also be found
in terms of combinations of $\mbox{erf}$ functions\cite{Houchmandzadeh2003},
but is not needed here. 

\begin{figure}
\begin{centering}
\includegraphics[width=0.9\columnwidth]{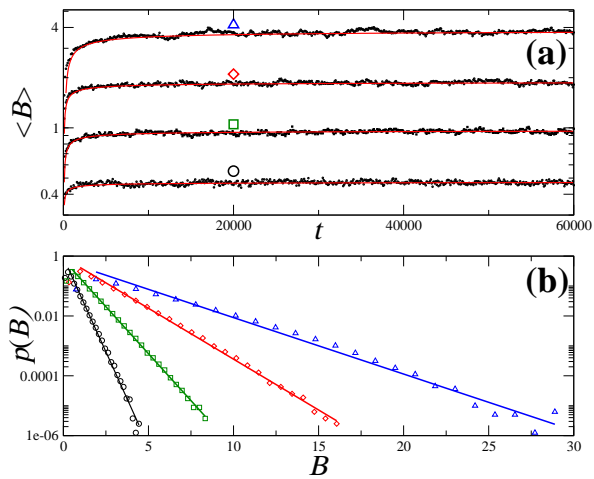}
\par\end{centering}
\caption{Front's width (eq. \ref{eq:width}) computed from numerical simulations
of the master equation (\ref{eq:MasterEquation}) by a Gillespie algorithm
comprising $M=2000$ sites (islands) for four sets of parameters $(N,m)$:
Black circles $(20,0.05)$ ($B_{\mbox{eq}}=0.475$); green squares
$(40,0.05)$ ($B_{\mbox{eq}}=0.975$); red diamonds $(20,0.2)$ ($B_{\mbox{eq}}=1.9$);
blue triangles $(40,0.2)$ ($B_{\mbox{eq}}=3.9$). (a) average front
$\left\langle B(t)\right\rangle $ computed over $10^{3}$ stochastically
generated $B(t)$, for $t\in[0,10^{5}]$. Dotted black curves : numerical
simulations ; solid red curves : theoretical prediction $B(t)=-Z_{0}(t)$
(eq\ref{eq:Znt}). (b) probability distribution of the width $B$
after equilibrium has been reached ($t\in[1,100]\times10^{5}$, sampling
time$10^{2}$ ) for the same parameters as in panel a. Symbols: numerical
simulations ; solid curves : exponential fits of the data $p(B)=A\,\exp(-aB)$
after the peak of the distribution has been reached\label{fig:Frontwidth}}
\end{figure}

As $B(0)=0$ and $\left\langle U(t)\right\rangle =0$, eq.(\ref{eq:Bt})
implies that 

\[
\left\langle B(t)\right\rangle =-Z_{0}(t).
\]
The front therefore reaches a finite width 
\begin{equation}
B_{\mbox{eq}}=m(N-1)/2\label{eq:Beq}
\end{equation}
and the equilibrium value is reached as $1/\sqrt{mt}$. Figure \ref{fig:Frontwidth}a
shows the perfect agreement of these results with numerical simulations.
The above equilibrium value of the width is also in agreement with
the value found from the SFKPP equation\cite{Hallatschek2009} $4Db^{-1}$
if the amplitude of the noise term is interpreted as $b=4\mu\ell/N$.
As noted by Hallatschek and Korolev\cite{Hallatschek2009}, genetic
drift alone can maintain a finite front width at $s=0$ in one dimension.
Moreover, numerical simulations of the discrete model show that the
width distribution probability of the front has an exponential tail
(figure \ref{fig:Frontwidth}b) 

Note that the width $B$ defined above as 
\[
\left\langle B\right\rangle =\sum_{i}\left\langle b_{i}\right\rangle =\frac{1}{N^{2}}\sum_{i}\left\langle n_{i}(N-n_{i})\right\rangle 
\]
 weights the regions with populations $0<n_{i}<N$, but contains no
information about their spatial distribution. A spatially wide front
composed for example of alternating $n=0$ and $n=N$ islands will
have $B=0$. 

\begin{figure}
\begin{centering}
\includegraphics[width=0.75\columnwidth]{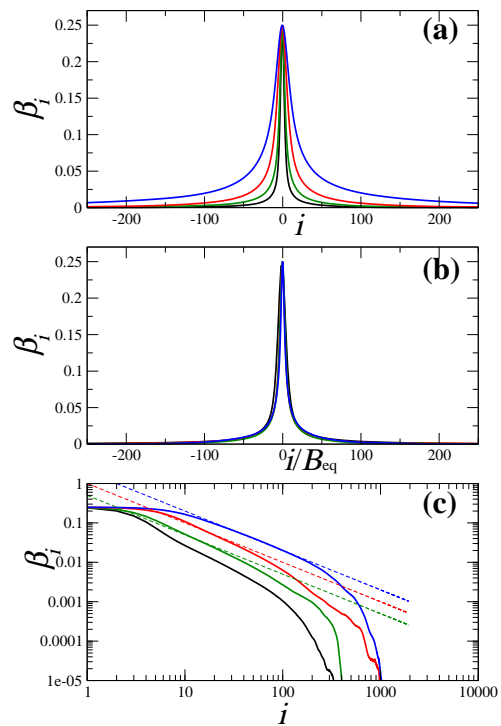}
\par\end{centering}
\caption{Mean front shape $\beta_{i}$ (eq. \ref{eq:betai}) as a function
of distance to the center of the front $i$, in the moving reference
frame. Numerical simulations schemes are the same as in figure \ref{fig:Frontwidth},
with $M=6000$. \label{fig:Mean-front-shape} In each sampled stochastic
realization, the displacement is computed from the relation (\ref{eq:displacement})
and the mutant population numbers in each site $i$ in a window of
1000-2000 sites around \emph{this} position are recorded. The mean
front shape $\nu_{i}$ (eq.\ref{eq:nui}) is computed on approximately
$10^{7}$ samples. (a) $\beta_{i}$ as a function of $i$ for various
$(N,m$) parameters : $(20,0.05)$ black, $(40,0.05)$ green, $(20,0.2)$
red, (40,0.2) blue. (b) Same as in panel (a), but the $x$ axis for
each curved is normalized by the corresponding equilibrium width $B_{\mbox{eq}}=(N-1)m/2$.
(c) The long tail of the front $\beta_{i}$ (solid curves), where
the parameters are the same as in panel (a). The dashed lines represent
the function $y=B_{\mbox{eq}}/2i$ as visual guides. }
 
\end{figure}

The shape of the front can be characterized more precisely by using
the moving frame of the front as the reference frame and computing
the mean relative mutant number $\nu_{i}$ and their weight $\beta_{i}$
in this frame 
\begin{eqnarray}
\nu_{i} & = & \frac{1}{N}\left\langle n_{i+[U]}\right\rangle \label{eq:nui}\\
\beta_{i} & = & \nu_{i}(1-\nu_{i})\label{eq:betai}
\end{eqnarray}
where $[U]$ is the integer part of the front displacement given by
the relation (\ref{eq:displacement}). These quantities are difficult
analytically but are readily computed by numerical simulation, as
shown in Figure \ref{fig:Mean-front-shape}. As it can be observed,
the mean front shape $\beta_{i}$ , which is a function of $N$ and
$m$ (Figure \ref{fig:Mean-front-shape}a) is spatially extended and
decreases slowly as a function of $i$, the distance to the center
of the front (Fig. \ref{fig:Mean-front-shape}c). The width $B_{\mbox{eq}}$
computed above remains however a good indicator of the mean front
shape, and all $\beta_{i}$ curves can be superimposed when the normalized
index $i/B_{\mbox{eq}}$ is used(Fig. \ref{fig:Mean-front-shape}b).

\begin{figure}
\begin{centering}
\includegraphics[width=0.9\columnwidth]{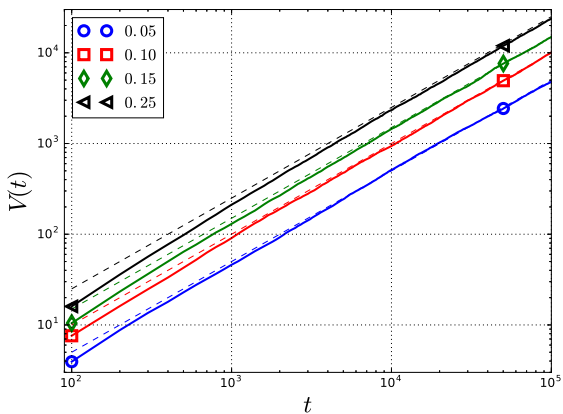}
\par\end{centering}
\caption{Solid curves: Variance of the position of the neutral front as a function
of time, computed from numerical simulations of the master equation
(\ref{eq:MasterEquation}) by a Gillespie algorithm, for various values
of the migration parameter $m$=0.05,0.1,0.15,0.25 and $N=10$. Dashed
lines: the theoretical values of $V(t)=mt$ (eq. \ref{eq:variance})
for corresponding $m$. The numerical simulations comprised $M=2000$
sites and the variance was computed over $6.4\times10^{3}$ stochastically
generated $U(t)$, for $t\in[0,10^{5}]$.\label{fig:variance}}
\end{figure}

The variance of the position of the front $V(t)=\left\langle U^{2}(t)\right\rangle -\left\langle U(t)\right\rangle ^{2}$
can be extracted by similar methods from equation (\ref{eq:vartiancedef})
: 
\begin{eqnarray*}
\frac{dV}{dt} & = & \frac{2}{N^{3}}\sum_{i}\left\{ \left\langle n_{i}(N-n_{i})\right\rangle -\frac{m}{2}\left\langle n_{i}n''_{i}\right\rangle \right\} \\
 & = & \frac{2}{N}\left\{ \left\langle B(t)\right\rangle -m\left(Z_{1}(t)-Z_{0}(t)\right)-\frac{m}{2}C_{0}\right\} .
\end{eqnarray*}
As $Z_{p}(t)$ can be computed exactly, the temporal evolution of
the variance can also be computed exactly. The result is particularly
simple for large times $t\gg1$
\begin{equation}
V(t)=mt+\mathcal{O}(\sqrt{mt})\label{eq:variance}
\end{equation}
The surprising result is that for large times, the diffusion of the
front is \emph{independent} of its width. The figure \ref{fig:variance}
shows the agreement of this expression with numerical solutions.

\subsection{Behavior of the front at small $s$.}

For a non-zero excess relative fitness, the moment closure does not
hold and the front characteristics can no longer be derived exactly.
It is however possible to derive the front speed to the first order
of the perturbation $s$. 

\begin{figure}
\begin{centering}
\includegraphics[width=0.95\columnwidth]{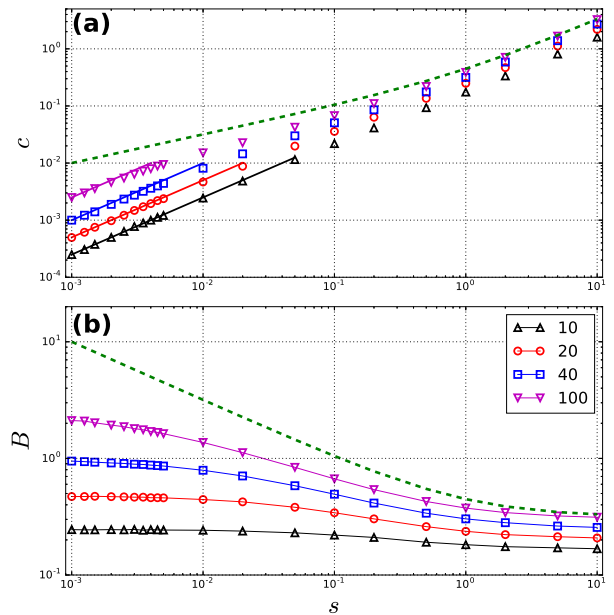}
\par\end{centering}
\caption{Speed $c$ (panel (a) ) and width $B$ (panel B) of the front as a
function of excess relative fitness $s$ for $m=0.05$ and $N=10$,
$20$,$40$, and $100$. Solid lines in panel (a) represent the first
order theoretical expression $c=Nms/2$ for $Ns\lesssim0.5$. Dashed
green lines represent the mean field values $c_{\mbox{m.f}}=\sqrt{2ms(s+1)}$
and $B_{\mbox{m.f}}=\sqrt{2m(s+1)/s}$. \label{fig:speed-s}}
\end{figure}

For $s>0$ the position of the front is given by 
\begin{eqnarray*}
\frac{d\left\langle U\right\rangle }{dt} & = & \frac{1}{N}\sum_{i}\left\langle W_{i}^{+}(\mathbf{n})-W_{i}^{-}(\mathbf{n})\right\rangle \\
 & = & s\left\{ B(t)-m\left(Z_{1}(t)-Z_{0}(t)\right)-\frac{m}{2}C_{0}\right\} 
\end{eqnarray*}
At small selection pressures $Ns\ll1$, on expanding the above expression
to the first order of perturbation, we find, in the limit of large
time 
\begin{equation}
\frac{d\left\langle U\right\rangle }{dt}=\frac{mNs}{2}+{\cal O}(1/\sqrt{t})\label{eq:dUdt}
\end{equation}
Note that at small $s$, the front speed scales as the selection \emph{pressure}
$Ns$. Even for $N=1$ when the front width $B_{eq}=0$, the front
acquires a non-zero speed (figure \ref{fig:speed-s}). 

The above computation of the position of the front at $s>0$, which
is a first order moment, requires the knowledge of second order moments
$Z_{i}$ at $s=0$. The same line of argument shows that computing
the variance of the position and width of the front for $s>0$ , which
are second order moments, necessitates the computation of third order
statistical quantities. Even though computation of higher momenta
is theoretically possible in the neutral case $s=0$, their effective
computation remains extremely tedious. 

Figure \ref{fig:speed-s} shows the result of stochastic based numerical
simulations for a wide range of $s$ and the agreement with expression
(\ref{eq:dUdt}) at small selection pressure. It can be observed that
the mean field approximation becomes correct only at very high excess
relative fitness $s$ and local population size $N$. Fluctuations
modify significantly the prediction of the FKPP model. 

\section{Microscopic model and general migration kernel.\label{sec:Microscopic model.}}

In the SFKPP description of the mutant wave (eq.\ref{eq:FKPP}) the
diffusion coefficient $D$ and the amplitude of the noise $b$ are
considered independent parameters. The same is true for the island
model of the preceding section where the population size of the island
$N$ and the migration probability $m$ between neighboring islands
were considered independent. However, at the individual level of evolution,
both migration and genetic drift are the result of the same phenomenon
of individuals replacing each other. These two parameters must therefore
be linked through an Einstein-like relation and cannot be independent. 

There is a level of arbitrariness in the island model in the manner
in which individuals are grouped together and\emph{ }deme size $N$
is chosen. As the amount of fluctuations is critically controlled
by $N$, the grouping process is crucial. This arbitrariness also
impacts on the migration probability. The very existence of a unique
migration rate between nearest neighbor islands can be brought into
question. Consider for example the low migration limit ($m\ll1$)
of the island model: two individuals physically far apart from each
other but grouped into the same island will have a higher probability
of replacing each other than two close individuals belonging to neighboring
islands. 
\begin{figure}
\begin{centering}
\includegraphics[width=0.85\columnwidth]{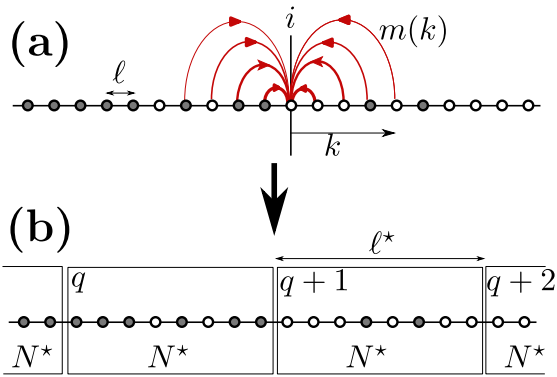}
\par\end{centering}
\caption{Neutral spatial Moran model with general migration kernel. (a) Individuals
are uniformly distributed in space ; when an individual at site $i$
dies, it is replaced by the progeny of its $k-$th neighbor with probability
$m(k)$. (b) Individuals are grouped into demes of population size
$N^{\star}$. In the classical scheme of SFKPP (figure \ref{fig:moranspace}),
the detailed migration kernel $m(k)$ is replaced by a single number
$m^{\star}$ of migration between neighboring demes. \label{fig:Population_continuous}}
\end{figure}

A more rigorous approach to this problem would be to consider a ``microscopic''
model where individuals are uniformly distributed in space and not
arbitrarily grouped into demes/islands. In this microscopic model,
migration/replacement is not restricted to nearest neighbors (figure
\ref{fig:Population_continuous}a): when an individual dies at site
$i$, it has a probability $m_{i}^{j}$ of being replaced by the progeny
of an individual at site $j$. Solving exactly this strictly individual
model then leads us to choose the \emph{effective }population size
of each deme, which is used in the stepping stone approach (figure
\ref{fig:Population_continuous}b) and derive the exact relation between
the diffusion coefficient and the noise amplitude, both of which are
a function of the dispersal kernel $m_{i}^{j}$. 

In this evolutionary graph approach, each site contains exactly one
individual (either wild type or mutant) $n_{i}=0,1$ ; the transition
probability densities for the number of mutants on site $i$ to increase/decrease
by one individual is a simple generalization of equations (\ref{eq:rateup},\ref{eq:ratedown})
. Thus:
\begin{eqnarray}
W_{i}^{+}(\mathbf{n}) & = & (1+s)\mu(1-n_{i})\sum_{j}m_{i}^{j}n_{j}\label{eq:rateupgen}\\
W_{i}^{-}(\mathbf{n}) & = & \mu n_{i}\sum_{j}m_{i}^{j}(1-n_{j})\label{eq:ratedowngen}
\end{eqnarray}
where $m_{i}^{j}$ is the probability that the progeny of an individual
at site $j$ replaces an individual at site $i$. In the literature
of plants, the migration probability $m_{i}^{j}$  is known as the
dispersal kernel and can be measured precisely in the field\cite{Nathan2000}.
In the following, we will consider dispersal kernels that depend only
on the distance between two sites, $i.e.$, $m_{i}^{j}=m(|j-i|)$. 

\subsection{Mean field approximation. }

Following the same steps as in subsection \ref{subsec:Mean-field-approximation.},
it is straightforward to deduce the mean field approximation of the
corresponding master equation. For a migration probability that depends
only on the distance between two sites, the mean field approximation
is exactly the same as expression (\ref{eq:meanfield}), where $\ell$
here is the inter-individual distance (lattice size), $\mu$ the death
rate and the diffusion coefficient 
\begin{equation}
D=\frac{\mu\ell^{2}}{2}\sum_{k}k^{2}m(k)\label{eq:D_contpop}
\end{equation}
 is given in terms of mean dispersal distance. 

From now on and to avoid confusion, we will refer to all quantities
derived in the island approximation (the macroscopic view) of section
\ref{sec:The-1d-system} by the super script $^{\star}$ . The diffusion
coefficient of the mean field approximation derived in subsection
\ref{subsec:Mean-field-approximation.} (relation \ref{eq:meanfield}),
for example, is 
\begin{equation}
D^{\star}=\frac{\mu\ell^{\star2}}{2}m^{\star}\label{eq:Dstar}
\end{equation}
For a 1d system, the patch extension $\ell^{\star}=N^{\star}\ell$
(figure \ref{fig:Population_continuous}b) ; comparing expression
(\ref{eq:D_contpop}) and (\ref{eq:Dstar}) therefore leads to 
\begin{equation}
N^{\star2}m^{\star}=\sum_{k}k^{2}m(k)\label{eq:N*-m*-relation}
\end{equation}
We observe here that the deme size $N^{\star}$ and the migration
probability between demes $m^{\star}$ are indeed linked through equation
(\ref{eq:N*-m*-relation}). 

\subsection{Stochastic characterization of the neutral front at $s=0$.\label{subsec:group}}

The relation (\ref{eq:N*-m*-relation}) is not sufficient to determine
the \emph{effective} population size $N^{\star}$ of islands. To address
this issue, we need to solve exactly the full stochastic model. We
restrict the computation here to the neutral case $s=0$, the derivations
for $s\ne0$ following precisely the steps developed previously. 

The computational approach is similar to subsection \ref{subsec:Stochastic-characterization}.
As before, the displacement of the front is defined as 
\begin{equation}
U(t)=\sum_{i=-\infty}^{+\infty}\left[n_{i}(t)-n_{i}(0)\right]\label{eq:displacement-microscopic}
\end{equation}
and 
\[
\frac{d}{dt}\left\langle U(t)\right\rangle =\sum_{i}\left\langle W_{i}^{+}(\mathbf{n})-W_{i}^{-}(\mathbf{n})\right\rangle 
\]
Therefore, for the neutral front $s=0$, $d_{t}\left\langle U\right\rangle =0$
and $\left\langle U(t)\right\rangle =0$. 

The second order moments are also defined as before 
\begin{equation}
Z_{p}(t)=\sum_{i}\left\langle n_{i}(t)n_{i+p}(t)\right\rangle -n_{i}(0)n_{i+p}(0)\label{eq:Zp-b}
\end{equation}
and we note that $Z_{p}=Z_{-p}.$ The equations governing $Z_{p}$
for the rescaled time $t\leftarrow\mu t$ are 
\begin{eqnarray}
\frac{d}{dt}Z_{0} & = & -2Z_{0}\label{eq:dZ0-b}\\
\frac{d}{dt}Z_{p} & = & 2\sum_{k=-\infty}^{\infty}m(k)\left(Z_{p+k}-Z_{p}\right)+C_{p}\,\,\,\,p\ne0\label{eq:dZp-b}
\end{eqnarray}
where the $C_{p}$ are defined by the initial condition 
\[
C_{p}=2\sum_{k}m(k)\sum_{i}n_{i}(0)\left(n_{i+p+k}(0)-n_{i+p}(0)\right)
\]
Note that equation (\ref{eq:dZ0-b}) implies that $Z_{0}(t)=0.$ This
is due to the fact that $n_{i}^{2}=n_{i}$ and therefore $Z_{0}(t)=\left\langle U(t)\right\rangle $. 

For an initially sharp front 
\begin{equation}
n_{i}=1\,\,\,\mbox{if}\,\,i\le0;\,\,=0\,\,\mbox{otherwise}\label{eq:sharpfront-b}
\end{equation}
which will be used here, 
\begin{equation}
C_{p}=2\sum_{k>\left|p\right|}m(k)\left(\left|p\right|-k\right)\label{eq:Cp}
\end{equation}

\begin{figure}
\begin{centering}
\includegraphics[width=0.95\columnwidth]{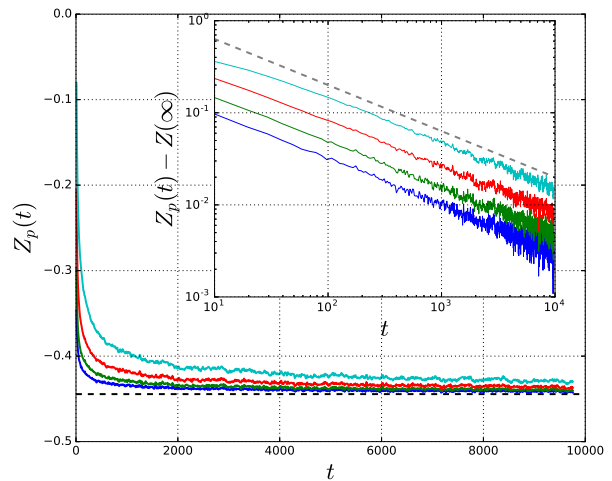}
\par\end{centering}
\caption{Numerical simulation of the second moment $Z_{p}$ (eq. \ref{eq:Zp-b})
in the neutral Moran model with geometric seed dispersal kernel (eq.
\ref{eq:dispersal-kernel}) for $p=1,2,4,8$. (Inset) Same data but
convergence to the limiting value $Z(\infty)$ is shown ; the gray
dashed line is $\propto1/\sqrt{t}$. Time is measured in units of
generation time ($1/\mu)$. For the numerical simulation, a lattice
of 4096 individuals is used and the result is averaged over $8\times10^{5}$
trials. \label{fig:Zp}}
\end{figure}

For simplicity, we further restrict the solution of equations (\ref{eq:dZ0-b}-\ref{eq:dZp-b})
to the generic geometric dispersal kernel 
\begin{equation}
m(k)=(1-\delta_{k,0})\frac{1-\lambda}{2\lambda}\lambda^{\left|k\right|}\label{eq:dispersal-kernel}
\end{equation}
where the parameter $\lambda$ controls the dispersal length $\kappa$:
\begin{equation}
\kappa^{2}=2\sum_{k>0}k^{2}m(k)=\frac{1+\lambda}{(1-\lambda)^{2}}\label{eq:kappa2}
\end{equation}
The case $N^{\star}=1$, $m^{\star}=1$ of the preceding section is
obtained when $\lambda\rightarrow0$. Note that in the framework of
the Moran model used here, a dead individual cannot replace itself,
hence $m(0)=0$. Moreover, for the geometric dispersal kernel, relation
(\ref{eq:Cp}) becomes
\[
C_{p}=\frac{-1}{1-\lambda}\lambda^{\left|p\right|}
\]
It is straightforward to check that again all $Z_{p}$ converge as
$\propto1/\sqrt{t}$ to the same value 
\begin{equation}
Z_{p}(\infty)=-\frac{\lambda}{(1-\lambda)^{2}}=\theta^{2}-\kappa^{2}\label{eq:Zp-mk}
\end{equation}
where $\kappa^{2}$ is defined in (\ref{eq:kappa2}) and $\theta=2\sum_{k>0}k\,m(k)$$=1/(1-\lambda)$.
We observe that the stationary value of $-Z_{p}$ is given by a quantity
similar to the variance of the dispersal kernel. Figure \ref{fig:Zp}
shows the agreement between these results and individual based numerical
simulation of the same system. 

\begin{figure}
\begin{centering}
\includegraphics[width=0.95\columnwidth]{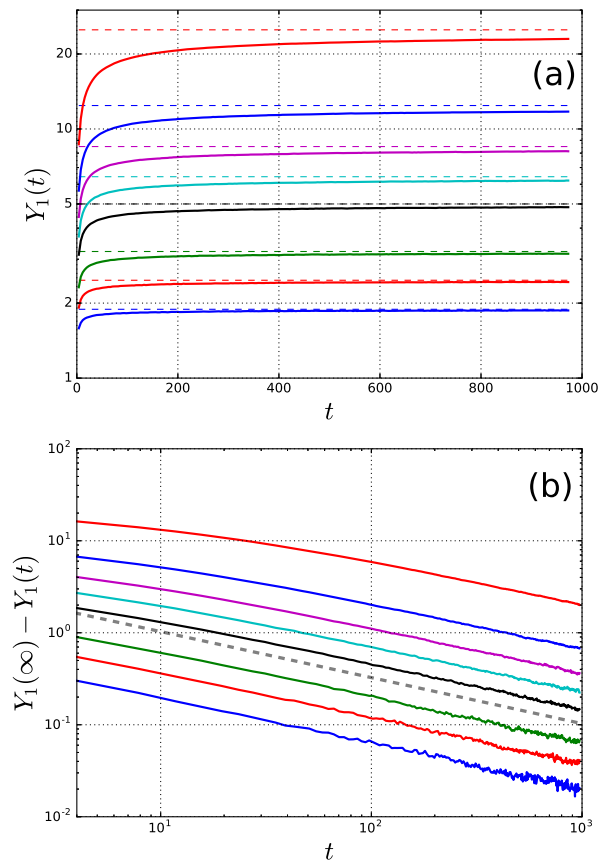}
\par\end{centering}
\caption{(a) Numerical simulation of the front width $Y_{1}(t)$ (eq. \ref{eq:Yp})
in the neutral Moran model with geometric seed dispersal kernel (eq.
\ref{eq:dispersal-kernel}) for increasing value of $\lambda=$0.25,0.33,0.4,0.5,0.55,0.6,0.66
and 0.75. (b) Same data but convergence to the limiting value $Y_{1}(\infty)$
is shown ; the gray dashed line is $\propto1/\sqrt{t}$. Time is measured
in units of generation time ($1/\mu)$. The numerical simulation parameters
are identical to figure \ref{fig:Zp}. \label{fig:Y1}}
\end{figure}

The width of the front can no longer be measured as in relation (\ref{eq:width})
by $B=\sum_{i}n_{i}(1-n_{i})$ which is always $0$. Other analog
metrics such as 
\begin{equation}
Y_{p}=\sum_{i}(n_{i}-n_{i+p})^{2}\label{eq:Yp}
\end{equation}
can be used to characterize the front. It is straightforward to show
that 
\begin{equation}
Y_{p}=p-2Z_{p}\label{eq:YpZp}
\end{equation}
Figure \ref{fig:Y1} shows the excellent agreement between the theoretical
results and the numerical simulations. 

The variance of the front displacement $V=\left\langle U^{2}\right\rangle -\left\langle U\right\rangle ^{2}$
can be computed by methods analogous to the previous section:
\begin{eqnarray*}
\frac{dV}{dt} & = & \sum_{i}\left(W_{i}^{+}+W_{i}^{-}\right)\\
 & = & -2\sum_{k}m(k)Z_{k}-C_{0}
\end{eqnarray*}
where the $W_{i}^{\pm}$ are the transition rates (\ref{eq:rateupgen},\ref{eq:ratedowngen})
and $C_{0}$ is defined by relation \ref{eq:Cp}). For the geometric
kernel $m(k)$ (eq. \ref{eq:dispersal-kernel}), using the long term
solution (\ref{eq:Zp-mk}) leads to 
\begin{equation}
V(t)=\kappa^{2}t+{\cal O}(\sqrt{t})\label{eq:Var-mk}
\end{equation}

\subsection{Grouping into islands.}

We now group individuals virtually into islands of size $N^{\star}$
and establish the condition under which the results obtained by the
SFKPP/islands model are valid. 

A patch of population size $N^{\star}$ is a virtual packing of individuals
into a deme, which we refer to by its index $q$ (Figure \ref{fig:Population_continuous}b).
The number of mutants in patch $q$ is 
\begin{equation}
n_{q}^{\star}=\sum_{a_{q}}^{b_{q}}n_{i}\label{eq:n*}
\end{equation}
where $b_{q}=a_{q}+N^{\star}-1$ and $a_{q+1}=b_{q}+1$. Note that
we have $N^{\star}$ possible choices for grouping individuals, as
we can set $a_{q}=N^{\star}q+r$, where $r=0$, 1, ...,$N^{\star}-1$.
We define the displacement and the width of the (macroscopic) front
as in subsection \ref{subsec:Stochastic-characterization} (definitions
\ref{eq:displacement},\ref{eq:width}) 
\begin{eqnarray}
U^{\star}(t) & = & \frac{1}{N^{\star}}\sum_{q=-\infty}^{+\infty}\left[n_{q}^{\star}(t)-n_{q}^{\star}(0)\right]\label{eq:displacement-group}\\
B^{\star}(t) & = & \frac{1}{N^{\star2}}\sum_{q=-\infty}^{+\infty}n_{q}^{\star}(t)\left[N^{\star}-n_{q}^{\star}(t)\right]\label{eq:width-group}
\end{eqnarray}
We compute the statistical properties of these quantities by taking
into account the detailed migration kernel (subsection \ref{subsec:group})
and compare them to the results obtained in subsection \ref{subsec:Stochastic-characterization}
where migrations were approximated by a single migration probability
$m^{\star}$ between neighboring demes. 

The macroscopic displacement $U^{\star}$ (relation \ref{eq:displacement-group})
is easily related to the microscopic displacement $U$ (relation \ref{eq:displacement-microscopic})
: 
\begin{eqnarray*}
U^{\star} & = & \frac{1}{N^{\star}}\sum_{q=-\infty}^{+\infty}\sum_{a_{q}}^{b_{q}}\left[n_{i}(t)-n_{i}(0)\right]\\
 & = & \frac{1}{N^{\star}}\sum_{i=-\infty}^{\infty}\left[n_{i}(t)-n_{i}(0)\right]\\
 & = & \frac{1}{N^{\star}}U
\end{eqnarray*}
and therefore $V^{\star}=\text{Var}(U^{\star})=\text{Var}(U)/N^{\star2}$.
The variance of the microscopic displacement $V$ is given by relation
(\ref{eq:Var-mk}), and hence, for long times, 
\begin{equation}
V^{*}=\frac{\kappa^{2}}{N^{\star2}}t\label{eq:Var-grouped}
\end{equation}
Comparing this expression to the relation $V^{\star}=m^{\star}t$
of the island model (section \ref{sec:The-1d-system}) where $m^{\star}$
is the migration probability between demes, we see that we must have
\begin{equation}
m^{\star}=\kappa^{2}/N^{\star2}\label{eq:mstar}
\end{equation}
The above relation is a confirmation of relation (\ref{eq:N*-m*-relation}),
which we obtained by a mean field approximation. We can also compute,
at small selection pressure $Ns$, the speed of the front that we
find to be $c=s\kappa^{2}/2$. Comparing this result to the speed
$c^{\star}=c/N^{\star}=sN^{\star}m^{\star}/2$ (eq. \ref{eq:dUdt})
of the island model leads again to the same relation between $m^{\star}$
and $N^{\star}$ as relation (\ref{eq:mstar}).

The macroscopic width of the front (relation \ref{eq:width-group})
can also be computed in terms of microscopic quantities (see Appendix
\ref{subsec:Relation-between-microscopic})
\[
B^{\star}=\frac{1}{N^{\star3}}\sum_{p=1}^{N-1}(N^{\star}-p)Y_{p}
\]
where $Y_{p}$ are the microscopic front width (defined by relation
\ref{eq:Yp}). For large times, relations (\ref{eq:YpZp}) and (\ref{eq:Zp-mk})
lead to 
\begin{equation}
B^{\star}=\frac{N^{\star}-1}{N^{\star2}}\left[\kappa^{2}-\theta^{2}+\frac{1}{6}(N^{\star}+1)\right]\label{eq:B*}
\end{equation}
The above relation is in perfect agreement with numerical simulations.
Comparing the above relation to the width of the front $B^{\star}=m^{\star}(N^{\star}-1)/2$
of the island model (relation \ref{eq:Beq}) and using relation (\ref{eq:mstar})
for $m^{\star}$, we find that we must have 
\begin{equation}
N^{\star}=3\theta-1\label{eq:N*}
\end{equation}
The above result determines the effective population size in terms
of the dispersal kernel. More precisely, 
\begin{eqnarray*}
N^{\star} & = & \frac{2+\lambda}{1-\lambda}\\
m^{*} & = & \frac{1+\lambda}{(2+\lambda)^{2}}
\end{eqnarray*}
We note that $m^{\star}$ is weakly dependent on $\lambda$ over its
whole range of variation {[}0, 1{]}, whereas $N^{\star}$ diverges
as the dispersion characteristic length when $\lambda\rightarrow1$. 

\section{Discussion and conclusion.}

In this article, we have used the formalism of the spatial Moran model
to study the propagation of mutants in a one dimensional geographically
extended population. The propagation of the mutant wave has usually
been studied in the framework of the SFKPP equation (\ref{eq:FKPP}).
The SFKPP equation however is a phenomenological approach and its
derivation from the fundamental models of population genetics such
as Wright-Fisher or Moran is not obvious. The deterministic Fisher
equation for the dynamics of the proportion $u(t)$ of a mutant in
a non-structured population is $d_{t}u=su(1-u)$. For geographically
structured populations, it seemed natural\cite{Fisher1937} to add
a spatial diffusion term and generalize simply this equation to $\partial_{t}u=D\Delta u+su(1-u)$,
where $u(x,t)$ is the \emph{local }proportion of the mutant. On the
other hand, since the time of Fisher and Wright, it was obvious that
genetic drift is an important factor at small selection pressure.
For \emph{large} non-structured ($d=0)$ populations, Kimura tackled
this problem by using a Fokker-Planck approximation of the Master
equations governing the WF or Moran models. The stochastic differential
equation associated with the Kimura equation is $d_{t}u=su(1-u)+\sqrt{u(1-u)/N}\eta(t)$.
It then seemed natural to unite the two approaches and propose the
SFKPP equation (\ref{eq:FKPP}). 

We see here that many assumptions were made in this process : (i)
the form of the diffusion term may be different ; (ii) $u(x,t)$ is
a local relative density and the noise term of SFKPP would be a good
approximation only if the number of individuals in each patch where
$u$ has been computed is large enough ; (iii) the noise term $\sqrt{u(1-u)/N}$
itself was obtained for a non-structured population and it is far
from obvious that it should be the same for an extended population
and not involve the spatial derivative of $u$. 

The individual based approached that we develop in this article is
intended to overcome these problems and to ground the SFKPP approach
on a firmer basis. Using an explicit spatial island model, we have
shown first that the diffusion term is indeed different from the FKPP
equation (relation \ref{eq:meanfield}) and this difference has important
consequences on the speed and width of the front for large selection
pressures (relation \ref{eq:speed_meanfield},\ref{eq:meanfield-width}). 

For small selection pressures, we derive the parameters of the front
(speed, diffusion coefficient and width) without any assumption on
the size of the local population and without neglecting the non-local
noise. These results are in agreement with the predictions of SFKPP
equation at small selection pressure as developed by \cite{Hallatschek2011}. 

Finally, by taking into account the explicit migration kernel, we
establish the relation between the amplitude of the diffusion and
that of the noise ; this approach also allows us to define the effective
size of the local population $N$, which is the crucial parameter
controlling the noise as a function of the dispersal length. 

Individual based models have the same level of complexity as their
equivalent stochastic differential equation approach. We believe that
the formalism developed in this article is a step forward in the search
for a better understanding of natural populations and the dynamics
of mutant waves. 
\begin{acknowledgments}
We are grateful to Erik Geissler for the critical reading of the manuscript
and fruitful discussions. 
\end{acknowledgments}

\section{Mathematical details.}

\subsection{The noise term in SFKPP\label{subsec:NoiseSFKPP}}

The argument for the phenomenological noise term used by Doering et
al. can be rephrased as follows in the framework of population genetics.
For a non-structured population (a population at $d=0)$, in an ecosystem
with carrying capacity of $N$ individuals formed of wild type individuals
with fitness $1$ and mutants with fitness $1+s$, the transition
rates for the one-step Moran process is \cite{Houchmandzadeh2010}
\[
W(n\rightarrow m)=\delta_{|m-n|,1}\frac{\mu}{N}(N-n)n(1+\delta_{m-n,1}s)
\]
where $n$ is the number of mutants. The probability $P(n,t)$ of
observing $n$ mutants at time $t$ is governed by the Master equation
associated with these rates 
\[
\frac{\partial P(n,t)}{\partial t}=\sum_{m}W(m\rightarrow n)P(m,t)-W(n\rightarrow m)P(n,t)
\]
For a \emph{large} ecosystem ($N\gg1$) at small selection pressure
$(Ns\ll1$), the above Master equation can be approximated by a Fokker-Plank
equation called the Kimura equation\cite{Kimura1962,Ewens2004,Houchmandzadeh2010}
\[
\frac{\partial P(u,t)}{\partial t}=-s\frac{\partial\left[u(1-u)P\right]}{\partial u}+\frac{1}{N}\frac{\partial^{2}\left[u(1-u)P\right]}{\partial u^{2}}
\]
where $u=n/N$ and time is measured in units of $1/\mu$. The above
diffusion equation is equivalent to a stochastic differential equation
for the density $u$\cite{Gardiner2004}:
\begin{equation}
\frac{du}{dt}=su(1-u)+\sqrt{u(1-u)/N}\eta(x,t)\label{eq:stodif-1}
\end{equation}
The origin of the noise is the genetic drift due to the size of the
system. 

For a spatially extended system, the Doering et al. phenomenological
approach to derive the SFKPP consists in adding a (spatial) diffusion
term $D\nabla^{2}u$ to the stochastic equation \ref{eq:stodif-1},
but conserving the same \emph{local} noise term, and neglecting fluctuations
from adjacent cells.

\subsection{Moment computation algebra.\label{subsec:Moment-computation-algebra.}}

The rules of moment computations are fairly standard (see for example
\cite{Pechenik1999}), but we give them here for self-consistency.
Various moments can be extracted directly from the Master Equation
\begin{eqnarray}
\frac{\partial P(\mathbf{n},t)}{\partial t} & = & \sum_{k}\left\{ W^{+}(a_{k}\mathbf{n})P(a_{k}\mathbf{n})-W^{+}(\mathbf{n})P(\mathbf{n})\right\} \nonumber \\
 & + & \sum_{k}\left\{ W^{-}(a_{k}^{\dagger}\mathbf{n})P(a_{k}^{\dagger}\mathbf{n})-W^{-}(\mathbf{n})P(\mathbf{n})\right\} \label{eq:master}
\end{eqnarray}
by multiplying it by some operator and then making the change of variable
$\mathbf{n}\rightarrow a_{k}\mathbf{n}$ or $a_{k}^{\dagger}\mathbf{n}$.
Consider for example 
\[
\frac{d\left\langle n_{i}n_{j}\right\rangle }{dt}=\frac{\partial}{\partial t}\sum_{\left\{ \mathbf{n}\right\} }n_{i}n_{j}P(\mathbf{n},t)
\]
After replacing $\partial_{t}P$ by its value from equation \ref{eq:master},
the first term on the r.h.s. of the above equation reads: 
\[
I_{1}=\sum_{\mathbf{n}}\sum_{k}n_{i}n_{j}W_{k}^{+}(a_{k}\mathbf{n})P(a_{k}\mathbf{n})-(...)
\]
Changing the variable $\mathbf{n}\rightarrow a_{k}^{\dagger}\mathbf{n}$
implies $n_{i}\rightarrow n_{i}+\delta_{i,k}$ and $n_{j}\rightarrow n_{j}+\delta_{j,k}$and
\begin{eqnarray*}
I_{1} & = & \sum_{\mathbf{n}}\sum_{k}\left(n_{i}+\delta_{i,k}\right)\left(n_{j}+\delta_{j,k}\right)W_{k}^{+}(\mathbf{n})P(\mathbf{n})-(...)\\
 & = & \left\langle n_{i}W_{j}^{+}+n_{j}W_{i}^{+}+\delta_{i,j}W_{i}^{+}\right\rangle 
\end{eqnarray*}
Computing now the second term and grouping all the terms, we get
\[
\frac{d\left\langle n_{i}n_{j}\right\rangle }{dt}=\left\langle n_{i}\left(W_{j}^{+}-W_{j}^{-}\right)\right\rangle +\left\langle n_{j}\left(W_{i}^{+}-W_{i}^{-}\right)\right\rangle +\delta_{i,j}\left\langle W_{i}^{+}+W_{i}^{-}\right\rangle 
\]
For the neutral case $s=0$ we have 
\[
W_{i}^{+}-W_{i}^{-}=\frac{m}{2}n''_{i}
\]
 and 
\[
W_{i}^{+}+W_{i}^{-}=\frac{2}{N}n_{i}(N-n_{i})+\frac{m}{2}n''_{i}-\frac{m}{N}n_{i}n''_{i}
\]
So finally 
\begin{eqnarray*}
\frac{d\left\langle n_{i}n_{j}\right\rangle }{dt} & = & \frac{m}{2}\left\langle n_{i}n''_{j}+n_{j}n''_{i}\right\rangle \\
 & + & \delta_{i,j}\left\{ \frac{2}{N}\left\langle n_{i}(N-n_{i})\right\rangle +\frac{m}{2}\left\langle n''_{i}\right\rangle -\frac{m}{N}\left\langle n_{i}n''_{i}\right\rangle \right\} 
\end{eqnarray*}

\subsection{Front speed and width in the mean field approximation.\label{subsec:meanfieldspeed}}

The minimum speed of propagating front of the equation
\begin{equation}
\frac{\partial u}{\partial t}=D\left[1+s(1-u)\right]\frac{\partial^{2}u}{\partial x^{2}}+\mu su(1-u)\label{eq:meanfield2}
\end{equation}
is obtained by following the original Fisher\cite{Fisher1937} approach.
For a propagating front, on setting $\partial_{t}u=-v\partial_{x}u$
and then setting $g=-du/dx$, we get the equation for $g(u)$ : 
\begin{equation}
D\left[1+s(1-u)\right]g\frac{dg}{du}-vg+\mu su(1-u)=0.\label{eq:gu}
\end{equation}
Setting $p=dg/du|_{u=0}$ as the slope of the curve at the origin
$u=0$, the equation for $p$ is 
\[
D(1+s)p^{2}-vp+\mu s=0
\]
which has a solution only for 
\[
v\ge v_{min}=2\sqrt{\mu Ds(1+s)}
\]

The width $B=\int_{\mathbb{R}}u(1-u)dx$ of the front can be computed
following the same approach. Setting $v=\partial_{t}\int_{\mathbb{R}}udx$,
exchanging the derivation on $t$ and integration on $x$, and performing
integration by parts on the propagating front, we get 
\begin{eqnarray*}
v & = & \mu sB+Ds\int_{\mathbb{R}}\left(\frac{du}{dx}\right)^{2}dx\\
 & = & \mu sB+Ds\int_{0}^{1}gdu
\end{eqnarray*}
The shape of the front $g(u)$ is not known. However, $g(u)$ is a
smooth function, $g(0)=g(1)=0$ and its slope at both ends, $p=dg/du|_{u=0}$
and $q=dg/du|_{u=1}$ are known and determined from equation (\ref{eq:gu}).
Approximating then $g(u)$ by a third order polynomial that respects
these constraints
\[
g(u)=u(1-u)(p-(p+q)u)
\]
leads to:
\begin{equation}
B=2\sqrt{\frac{D(1+s)}{\mu s}}\kappa(s)\label{eq:meanwidth}
\end{equation}
where 
\begin{eqnarray*}
\kappa(s) & = & 1-\frac{s}{24(1+s)}\left(\left(\sqrt{\frac{2+s}{1+s}}-1\right)(s+1)+1\right)\\
 & = & \frac{15}{16}+\frac{13}{192s}+{\cal O}(s^{-2})\,\,\,\,\mbox{for}\,\,s\gg1
\end{eqnarray*}

\subsection{Choice of initial conditions.\label{subsec:Choice-of-initial}}

The definition of various moments we use in this article such as relations
(\ref{eq:displacement}-\ref{eq:Zp}) ensures that the infinite sums
over sites contain only a finite number of non-zero terms; it avoids
the problem of spurious effects due to manipulation of divergent series.
However, the initial front may not need to be sharp, but only finite. 

Consider the discrete function 
\[
f_{i}=N\,\,\,\mbox{if}\,\,i\le0;\,\,=0\,\,\mbox{otherwise}
\]
where the position $0$ corresponds the to middle of the initial front.
We can redefine the moments as 
\begin{eqnarray*}
U & = & \sum_{i}n_{i}(t)-f_{i}\\
Z_{p} & = & \frac{1}{N^{2}}\sum_{i}\left\{ \left\langle n_{i}(t)n_{i+p}(t)\right\rangle -f_{i}f_{i+p}\right\} 
\end{eqnarray*}
The differential equations we derived throughout this article remain
invariant under this definition of the moments, the only difference
being that the initial values of these moments are non-zero. 

\subsection{Relation between microscopic and macroscopic front width.\label{subsec:Relation-between-microscopic}}

By definition, 
\begin{eqnarray*}
N^{\star2}B^{\star} & = & \sum_{q}n_{q}^{\star}(N^{\star}-n_{q}^{\star})\\
 & = & \sum_{q}\sum_{i,j=a_{q}}^{b_{q}}n_{i}(1-n_{j})\\
 & = & \sum_{q}\left\{ \sum_{i=a_{q}}^{b_{q}}n_{i}(1-n_{i})+\sum_{i=a_{q}}^{b_{q}-1}\sum_{j=i+1}^{b_{q}}(n_{i}+n_{j}-2n_{i}n_{j})\right\} 
\end{eqnarray*}
where $n^{\star}$is defined by relation (\ref{eq:n*}), $b_{q}=a_{q}+N^{\star}-1$
and $a_{q+1}=b_{q}+1$. Note that we have $N^{\star}$ possible choices
for grouping individuals, as we can set $a_{q}=N^{\star}q+r$, where
$r=0$, 1, ...,$N^{\star}-1$. As $n_{i}=0,1,$ the first term in
the above sum is zero and $n_{i}^{2}=n_{i}$. Rearranging the indices
in the second term, we have 
\begin{equation}
N^{\star2}B^{\star}=\sum_{q}\sum_{k=1}^{N^{\star}-1}\sum_{i=a_{q}}^{b_{q}-k}(n_{i}+n_{i+k}-2n_{i}n_{i+k})\label{eq:seq}
\end{equation}

Note that for an unrestricted sum 
\[
\sum_{i=-\infty}^{\infty}n_{i}+n_{i+p}-2n_{i}n_{i+p}=Y_{p}
\]
However, the problem with expression (\ref{eq:seq}) is that we are
missing the terms $n_{i}n_{i+k}$ , where $i$ is in one cell and
$i+k$ in another one. In fact, for each cell, we are missing $k$
terms of the form $n_{i}n_{i+k}$ connecting two neighboring cells.
We now use our freedom to choose the grouping $r$: we use $N^{\star}$
different choices of $r$ and sum all of them. A term missing in one
choice of $r$ will be recovered in another. As the result must not
depend on the choice of $r$, we have 
\begin{eqnarray*}
N^{\star}(N^{\star2}B^{\star}) & = & \sum_{p=1}^{N^{\star}-1}(N^{\star}-p)Y_{p}
\end{eqnarray*}
The index manipulation is clearer when performed manually on a few
simple examples such as $N^{\star}=2$ or 3. 

\subsection{Numerical simulations.}

All numerical simulations are written in C++, and data analysis is
performed by the high level language Julia\cite{Bezanson2014}. Numerical
simulation of the island model (section \ref{sec:The-1d-system})
is performed by a Gillespie algorithm by computing the jump probabilities
from the transition rates (\ref{eq:rateup},\ref{eq:ratedown}). For
the generalized migration kernel of section \ref{sec:Microscopic model.},
the Gillespie approach is too cumbersome and a direct approach has
been used: the index of an individual is chosen at random and it is
replaced by the value of another individual chosen according to the
kernel $m(k)$. 

\bibliographystyle{unsrt}

\begin{thebibliography}{}

\end{thebibliography}


\begin{thebibliography}{10}

\bibitem{Kimura1962}
M~Kimura.
\newblock {On the probability of fixation of mutant genes in a population.}
\newblock {\em Genetics}, 47:713--719, 1962.

\bibitem{Houchmandzadeh2010}
B~Houchmandzadeh and M~Vallade.
\newblock {Alternative to the diffusion equation in population genetics.}
\newblock {\em Phys Rev E Stat Nonlin Soft Matter Phys}, 82(5 Pt 1):51913,
  2010.

\bibitem{Lieberman2005}
Erez Lieberman, Christoph Hauert, and Martin~A Nowak.
\newblock {Evolutionary dynamics on graphs.}
\newblock {\em Nature}, 433(7023):312--316, 2005.

\bibitem{Houchmandzadeh2011b}
Bahram Houchmandzadeh and Marcel Vallade.
\newblock {The fixation probability of a beneficial mutation in a
  geographically structured population}.
\newblock {\em New Journal of Physics}, 13(7):073020, jul 2011.

\bibitem{Fisher1937}
R.A. Fisher.
\newblock {The wave of advance of advantageous genes.}
\newblock {\em Annals of Eugenics}, 7:355--369, 1937.

\bibitem{Kolmogorov}
A.~N. Kolmogorov.
\newblock {A study of diffusion equation with increase in the amount of
  substance}.
\newblock In {\em Selected work of N.A. Kolmogorov}, chapter~38, pages
  242--271. Kluwer Academic Publishers, 1991.

\bibitem{Murray2007}
James~D. Murray.
\newblock {\em {Mathematical Biology: I. An Introduction (Interdisciplinary
  Applied Mathematics) (Pt. 1)}}.
\newblock Springer, 2007.

\bibitem{Beuer1995}
H.P. Beuer, W.~Huber, and F.~Petruccione.
\newblock {The Macroscopic Limit in a Stochastic Reaction-Diffusion Process}.
\newblock {\em Europhysics Letter}, 30(April):69--74, 1995.

\bibitem{Munier2003}
S.~Munier and R.~Peschanski.
\newblock {Geometric Scaling as Traveling Waves}.
\newblock {\em Physical Review Letters}, 91(23):232001, dec 2003.

\bibitem{Vansaarloos2003}
W~Vansaarloos.
\newblock {Front propagation into unstable states}.
\newblock {\em Physics Reports}, 386(2-6):29--222, nov 2003.

\bibitem{Kingsolver2001}
J~G Kingsolver, H~E Hoekstra, J~M Hoekstra, D~Berrigan, S~N Vignieri, C~E Hill,
  a~Hoang, P~Gibert, and P~Beerli.
\newblock {The strength of phenotypic selection in natural populations.}
\newblock {\em The American naturalist}, 157(3):245--61, mar 2001.

\bibitem{Nei2005}
Masatoshi Nei.
\newblock {Selectionism and neutralism in molecular evolution.}
\newblock {\em Molecular biology and evolution}, 22(12):2318--42, dec 2005.

\bibitem{Brunet1997}
Eric Brunet and Bernard Derrida.
\newblock {Shift in the velocity of a front due to a cutoff}.
\newblock {\em Physical Review E}, 56(3):2597--2604, sep 1997.

\bibitem{Doering2003}
Charles~R. Doering, Carl Mueller, and Peter Smereka.
\newblock {Interacting particles, the stochastic
  Fisher-Kolmogorov-Petrovsky-Piscounov equation, and duality}.
\newblock {\em Physica A: Statistical Mechanics and its Applications},
  325(1-2):243--259, jul 2003.

\bibitem{Hallatschek2009}
Oskar Hallatschek and K.~Korolev.
\newblock {Fisher Waves in the Strong Noise Limit}.
\newblock {\em Physical Review Letters}, 103(10):108103, sep 2009.

\bibitem{Ethier1977}
S.~N. Ethier and M.~F. Norman.
\newblock {Error estimate for the diffusion approximation of the Wright--Fisher
  model.}
\newblock {\em Proc Nat Acad Sci USA}, 74:5096--5098, 1977.

\bibitem{Korolev2010}
K~S Korolev, Mikkel Avlund, Oskar Hallatschek, and David~R Nelson.
\newblock {Genetic demixing and evolution in linear stepping stone models.}
\newblock {\em Reviews of modern physics}, 82(2):1691--1718, jun 2010.

\bibitem{Ewens2004}
W~J Ewens.
\newblock {\em {Mathematical Population Genetics}}.
\newblock Springer-Verlag, 2004.

\bibitem{Fisher1999}
R~A Fisher.
\newblock {\em {The genetical theory of natural selection, a complete variorum
  edition}}.
\newblock Oxford University Press, 1999.

\bibitem{Moran1962}
P~Moran.
\newblock {\em {Statistical Processes of Evolutionary Theory}}.
\newblock Clarendon Press, 1962.

\bibitem{Kimura1964a}
M~Kimura and G~H Weiss.
\newblock {The Stepping Stone Model of Population Structure and the Decrease of
  Genetic Correlation with Distance.}
\newblock {\em Genetics}, 49(4):561--76, apr 1964.

\bibitem{Maruyama1974}
T~Maruyama.
\newblock {A Markov process of gene frequency change in a geographically
  structured population.}
\newblock {\em Genetics}, 76(2):367--377, 1974.

\bibitem{Hubbel2001}
Stephen~P. Hubbell.
\newblock {\em {The unified neutral theory of Biodiversity and Biogeography.}}
\newblock Princeton University Press, 2001.

\bibitem{Houchmandzadeh2003}
B~Houchmandzadeh and M~Vallade.
\newblock {Clustering in neutral ecology.}
\newblock {\em Phys Rev E Stat Nonlin Soft Matter Phys}, 68(6 Pt 1):61912,
  2003.

\bibitem{Vallade2003}
M~Vallade and B~Houchmandzadeh.
\newblock {Analytical solution of a neutral model of biodiversity.}
\newblock {\em Phys Rev E Stat Nonlin Soft Matter Phys}, 68:61902, 2003.

\bibitem{Panja2003}
Debabrata Panja.
\newblock {Asymptotic scaling of the diffusion coefficient of fluctuating
  pulled fronts}.
\newblock {\em Physical Review E}, 68(6):065202, dec 2003.

\bibitem{Nathan2000}
R~Nathan and Hc~Muller-Landau.
\newblock {Spatial patterns of seed dispersal, their determinants and
  consequences for recruitment.}
\newblock {\em Trends in ecology {\&} evolution}, 15(7):278--285, jul 2000.

\bibitem{Hallatschek2011}
Oskar Hallatschek.
\newblock {The noisy edge of traveling waves.}
\newblock {\em Proceedings of the National Academy of Sciences of the United
  States of America}, 108(5):1783--7, feb 2011.

\bibitem{Gardiner2004}
C~Gardiner.
\newblock {\em {Handbook of Stochastic Methods: for Physics, Chemistry and the
  Natural Sciences}}.
\newblock Springer, 2004.

\bibitem{Pechenik1999}
Leonid Pechenik and Herbert Levine.
\newblock {Interfacial velocity corrections due to multiplicative noise}.
\newblock {\em Physical Review E}, 59(4):3893--3900, apr 1999.

\bibitem{Bezanson2014}
Jeff Bezanson, Alan Edelman, Stefan Karpinski, and Viral~B. Shah.
\newblock {Julia: A Fresh Approach to Numerical Computing}.
\newblock {\em arxiv}, page 1411.1607, nov 2014.

\end{thebibliography}

\end{document}